\documentclass[aps,prd,superscriptaddress,nofootinbib,amsmath,amsfonts,preprintnumbers,notitlepage,10pt,english]{revtex4}
\usepackage{amsmath}
\usepackage{amssymb}
\usepackage{babel}
\usepackage{graphicx}
\usepackage{color}

\def\d{\delta}
\newcommand{\f}[2]{\frac{#1}{#2}}
\newcommand{\mk}[1]{\left( #1 \right)}
\newcommand{\kk}[1]{\left[ #1 \right]}

\newcommand{\be}{\begin{equation}}
\newcommand{\ee}{\end{equation}}
\newcommand{\bea}{\begin{eqnarray}}
\newcommand{\eea}{\end{eqnarray}}
\def\Mpl{M_{\rm Pl}}



\begin{document}

\title{Constraint on reheating after $f(R)$ inflation from gravitational waves}

\author{Atsushi Nishizawa}
\email{nishizawa@tap.scphys.kyoto-u.ac.jp}
\affiliation{Department of Physics, Kyoto University, Kyoto 606-8502, Japan}

\author{Hayato Motohashi}
\email{motohashi@kicp.uchicago.edu}
\affiliation{Kavli Institute for Cosmological Physics, The University of Chicago, 
Chicago, Illinois 60637, U.S.A. }

\begin{abstract}
Recently, a combined model of the primordial inflation and the present cosmic acceleration has been proposed in the context of $f(R)$ gravity. This model is composed of the late-time acceleration term and an $R^2$ term, which enables the model to avoid high curvature singularity and describe a quasi-de Sitter inflationary phase in the early Universe. An interesting feature of this model is that the reheating dynamics after the inflation is significantly modified, in contrast to the original $R^2$ model, and affects the shape of a gravitational wave background (GWB) spectrum. Here we investigate the production of a GWB during the inflation and reheating eras in the $R^2$-corrected $f(R)$ model and compute a GWB spectrum. We found that interesting region of the model parameters has already been excluded by the cosmological limit on abundance of GWs.
\end{abstract}

\date{\today}

\maketitle

\section{Introduction}
The physical origin of two accelerated expansion regimes of the early and late-time Universe has been veiled for a long time and is one of the most important issues to address for modern cosmology. Various theoretical models have been proposed to explain accelerated expansion (For a review, see e.g. \cite{Tsujikawa:2010sc} and references therein): cosmological constant, additional energy components of the Universe such as scalar fields (inflaton or dark energy), modification of gravity at large distance, and matter inhomogeneities. Although most of them describe one of the two accelerated expansions, it is actually possible to deal with both of them at the same time in the framework of $f(R)$ gravity.

$f(R)$ gravity is a fourth-order theory of gravity, which is relatively simple and nontrivial generalization of general relativity (for a recent review, see \cite{Sotiriou:2008rp,DeFelice:2010aj}). It generalizes the Einstein-Hilbert action by introducing a function of scalar curvature $f(R)$. By choosing suitable functional form of $f(R)$, one can describe accelerated expansion of the Universe because the additional degree of freedom of the function $f(R)$ plays a role of a scalar field, which is called scalaron, and is responsible for the acceleration. The original idea of $f(R)$ modification has been proposed in \cite{Starobinsky:1980te}, where de Sitter expansion was derived as a solution for the Einstein equation with quantum one-loop contributions. If one writes down the action for this Einstein equation in the presence of one-loop terms, the action includes $R^2$ term. It tells us that the modification of the Einstein-Hilbert action by adding $R^2$ term admits de Sitter expansion. In addition, this de Sitter expansion is followed by the gravitational reheating and subsequent radiation dominated era. Thus, it is a self-consistent scenario of the early Universe and is referred to as the $R^2$ inflation model. The prediction of the $R^2$ model is slightly red-tilted spectrum and modest tensor-to-scalar ratio, which are consistent with recent observational data \cite{Ade:2013zuv}.

Meanwhile, $f(R)$ gravity can also explain the late-time acceleration. After some early challenges, the viable $f(R)$ models were proposed which realize stable matter-dominated regime and subsequent late-time cosmic acceleration~\cite{Hu:2007nk,Appleby:2007vb,Starobinsky:2007hu}. In these models, the expansion history of the Universe is close to that in the concordance ${\rm \Lambda}$-cold dark matter (${\rm \Lambda}$CDM) model at the recent epoch. However, the viable $f(R)$ models for the late-time acceleration still have theoretical problems~\cite{Starobinsky:2007hu} such as divergence of the scalaron mass in the past~\cite{Tsujikawa:2007xu} and curvature singularity~\cite{Appleby:2008tv,Frolov:2008uf,Kobayashi:2008tq}. To cure the theory of these pathological behaviors, one needs some correction for high curvature regime. Actually, it has been found that $R^2$ correction works well~\cite{Appleby:2009uf}. As a result, the $R^2$-corrected $f(R)$ model ($gR^2$-AB model) has the late-time acceleration term and the $R^2$ term which drives the inflation. It is interesting that reheating after inflation in this model is significantly different from that in the pure $R^2$ model, which is the main theme of the previous work~\cite{Motohashi:2012tt} and the present paper.

In an observational side, it is important to distinguish small deviation of $f(R)$ models from the ${\rm \Lambda}$CDM model since $f(R)$ models are constructed so as to reproduce the cosmic expansion in the ${\rm \Lambda}$CDM model at background level. Since in $f(R)$ gravity the effective gravitational constant depends on time and distance scale, the growth of the matter density fluctuations is enhanced at cosmological distance and is useful to measure the deviation \cite{Zhang:2005vt,Tsujikawa:2007gd,Hu:2007nk,Starobinsky:2007hu,Tsujikawa:2009ku,Motohashi:2009qn,Narikawa:2009ux,Motohashi:2010tb}. Also, this enhancement mitigates cosmological constraint on neutrino mass, allowing its total mass up to $\sim 0.5\,{\rm{eV}}$ \cite{Motohashi:2010sj} and sterile neutrino mass up to $\sim 1\,{\rm{eV}}$ \cite{Motohashi:2012wc}, because massive neutrinos suppress the evolution of matter fluctuations by free streaming and cancels the anomalous enhancement of matter growth in $f(R)$ gravity. Other distinguishable features of $f(R)$ gravity would be imprinted on cosmological gravitational waves (GWs) \cite{Ananda:2007xh,Capozziello:2007vd,Alves:2009eg}. Future pulsar timing experiments and gravitational-wave detectors will be able to probe them directly and test gravity theories \cite{Lee:2008ApJ,Chamberlin:2011ev,Nishizawa:2009bf,Nishizawa:2009jh,Nishizawa:2013eqa}.

In this paper, we investigate the production of a gravitational wave background (GWB) during the inflation and reheating regimes in the $gR^2$-AB model \cite{Appleby:2009uf} and discuss the observational constraint on the model from GWs. In this model, the gravity action $f(R)$ is elaborated so as to smoothly connect two accelerated cosmic expansions in the early Universe and the present time, avoiding instability and singularity of the model.  The inflation and reheating dynamics in the model have already been studied in the Jordan frame \cite{Appleby:2009uf} and in the Einstein frame \cite{Motohashi:2012tt}. Inflation is driven by $R^2$ term and thus time evolution is the same as that in the original $R^2$ inflation model. However, reheating is quite different from the $R^2$ model because of an additional term in the $f(R)$ action. As a result, the modification of gravity alters cosmic expansion during the reheating phase, whose analytic solution is systematically derived in \cite{Motohashi:2012tt}, as if there exists effective fluid with the equation of state of $w\equiv p/\rho=1$. Thus we expect that a GWB spectrum at high frequencies is significantly enhanced. One might consider that it is easy to construct the other specific functional forms of $f(R)$ that avoid singularities and describe both the primordial and present cosmic accelerated expansions. However, these functions belong to the same class because the stability conditions demand them to have similar behaviors even though they have different parameterizations. In this sense, it is worth studying one specific model in detail as an example of such a class of extended $f(R)$ models.

This paper is organized as follows. In Sec.~\ref{sec-be}, we briefly review the basic equations in $f(R)$ gravity in both the Jordan frame and the Einstein frame. We also present the inflation and reheating dynamics in the $gR^2$-AB model and their analytical solutions. In Sec.~\ref{secGW}, we define effective cosmic expansion including the contribution of modified gravity and compute a GWB spectrum with quantum-field formulation. In Sec.~\ref{sec:constraint}, we derive the constraint on model parameters of the $R^2$-corrected $f(R)$ model from an observational limit on the abundance of GWs. Sec.~\ref{sec-cn} is devoted to conclusions and discussion. Throughout the paper, we adopt units $c=\hbar=1$.

\section{$f(R)$ inflation and reheating}
\label{sec-be}

We briefly review the basic equations of $f(R)$ gravity theory and a viable model, so-called the $gR^2$-AB model, which unifies the primordial inflation and the present accelerating-expansion of the Universe. Also we summarize the results of \cite{Motohashi:2012tt}, which are needed for the computation of a GWB spectrum after Sec.~\ref{secGW}.

\subsection{$f(R)$ gravity and conformal transformation}

$f(R)$ gravity is defined by the action
\be S=\int d^4x \sqrt{-g} \left[ \f{M_{\rm{Pl}}^2}{2} f(R) + {\cal L}_M(g_{\mu\nu}) \right] \;, \ee
where ${\cal L}_M$ is the Lagrangian density for the matter sector and $M_{\rm{Pl}}\equiv (8\pi G)^{-1/2}$ is the reduced Planck mass. 
In general, the equation of motion is fourth-order differential equation and is difficult to solve. To analyze the inflation and the reheating in $f(R)$ gravity, it is useful to perform the conformal transformation and move to the Einstein frame, in which the additional degree of freedom due to modified action of gravity is interpreted as a scalar field with a potential term. Then we have the second-order equation of motion and are able to use the analogy of single-field inflation. Since we regard the Jordan frame as the physical frame, we need to recast resultant quantities obtained in the Einstein frame back to the Jordan frame after the calculation. Transforming the metric as $\tilde{g}_{\mu\nu}=F(R) g_{\mu\nu}$ and defining the canonical scalar field $\phi$, the scalaron, as 
\be F(R)\equiv \frac{d\,f(R)}{dR} \equiv e^{\sqrt{\f{2}{3}}\f{\phi}{\Mpl}}, \label{FRJ} \ee
the action is rewritten as
\be S=\int d^4x \sqrt{-\tilde{g}} \kk{\f{M_{\rm{Pl}}^2}{2} \tilde{R}-\f{1}{2}\tilde{g}^{\mu\nu} \partial_\mu \phi \partial_\nu \phi-V(\phi) +{\cal L}_M \left(e^{-\sqrt{\f{2}{3}}\f{\phi}{\Mpl}} \tilde{g}_{\mu\nu} \right)}, \ee
with the potential term
\be V(\phi)=\frac{\Mpl^2}{2} \f{R(\phi)F(R(\phi))-f(R(\phi))}{F(R(\phi))^2}. \nonumber \ee
In these equations, the tildes denote physical quantities in the Einstein frame. The Einstein equation in the Einstein frame reduces to 
\bea
\tilde{H}^2&=&\f{1}{3\Mpl^2}\kk{\f{1}{2}\mk{\f{d\phi}{d\tilde{t}}}^2 + V(\phi) + \tilde{\rho}}, \label{Eeq1} \\
\f{d\tilde{H}}{d\tilde{t}}&=&-\f{1}{2\Mpl^2}\kk{\mk{\f{d\phi}{d\tilde{t}}}^2 + \tilde{\rho}+\tilde{P}}. \label{Eeq2}
\eea
The scalar field obeys the equation of motion,
\be \f{d^2\phi}{d\tilde{t}^2}+3\tilde{H}\f{d\phi}{d\tilde{t}}+V_{,\phi}(\phi)=\f{1}{\sqrt{6} \Mpl} (\tilde{\rho}-3\tilde{P}). \label{Eeq3} \ee

By the conformal transformation, the time coordinates and scale factors in both frames are related by
\be dt=e^{-\f{1}{\sqrt{6}}\f{\phi}{\Mpl}}d\tilde{t}, \quad a=e^{-\f{1}{\sqrt{6}}\f{\phi}{\Mpl}}\tilde{a}. \label{ta} \ee
From the above definitions, the transformation of the Hubble parameter is given by 
\be H=e^{\f{1}{\sqrt{6}}\f{\phi}{\Mpl}}\mk{\tilde{H}-\f{1}{\sqrt{6}\Mpl}\f{d\phi}{d\tilde{t}}}. \label{HJ} \ee

\subsection{$gR^2$-AB model}

The $gR^2$-AB model~\cite{Appleby:2009uf} has been proposed to combine the accelerated expansions in both the early and the present Universes, satisfying several conditions for theoretical stability during whole cosmological expansion history. The model is described by the following $f(R)$ action, 
\be f (R)=(1-g)R+g M^2\d \log\kk{\f{\cosh(R/M^2\d-b)}{\cosh b}}+\f{R^2}{6M^2}, \label{gR2AB} \ee
where $g$, $b$, $\d$, and $M$ are positive-definite model parameters. $g$ should be in the range of $0<g<1/2$ to hold the stability conditions of $f(R)$ gravity: $F(R)>0$ and $dF(R)/dR>0$, where
\be F(R)=1-g+\f{R}{3M^2}+g \tanh (R/M^2\d -b). \label{FgABR2} \ee
To make the physical role of each term more transparent, the above $f(R)$ function is equivalently expressed as
\begin{align}
f(R) &=R-\frac{R_{\rm{vac}}}{2} + g M^2 \delta \log \left[ 1+e^{-2(R/M^2\delta -b)} \right]+\frac{R^2}{6M^2} \;, 
\label{eq58} \\
R_{\rm{vac}} &\equiv 2 g M^2 \delta \left\{ b+\log(2\cosh b) \right\}. \nonumber 
\end{align}

In high curvature regime $R \gg M^2$, the fourth term in Eq.~(\ref{eq58}) dominates and causes nearly de Sitter inflationary expansion of the Universe, which is the same as the $R^2$ inflation~\cite{Starobinsky:1980te}. The parameter $M$ determines the energy scale of inflation. Since the e-folding number when the cosmic microwave background (CMB) scale today exits horizon during the inflation is $N\sim 66$ counted from the end of the inflation, the amplitude of the temperature fluctuation of CMB anisotropy at $k=0.002\,{\rm{Mpc}}^{-1}$ \cite{Hinshaw:2012fq} fixes the parameter $M$ to $M \approx 1.2 \times 10^{-5}\,M_{\rm{Pl}}$ \cite{Motohashi:2012tt}. For this choice of $M$, the spectral indices of the scalar and tensor modes and the tensor-to-scalar ratio defined at the CMB scale are given by $n_S-1 \approx -2/N \approx 0.97$, $n_T \approx -3/(2 N^2) \approx -3.4 \times 10^{-4}$, and $r \approx 12/N^2 \approx 2.8 \times 10^{-3}$, at the leading order in the slow-roll parameter \cite{Hwang:2001pu}. So the current observations \cite{Hinshaw:2012fq} are all consistent with the values of $n_S$, $n_T$, and $r$ predicted by the $gR^2$-AB model.

On the other hand, in low curvature regime, the first and second terms in Eq.~(\ref{eq58}) are equivalent of general relativity with small cosmological constant at present. Thus, the parameter $\delta$ should be determined so that the current observation of accelerated expansion, $R_{\rm{vac}} \sim 10^{-120} M_{\rm{Pl}}^2$, is reproduced, and is given by 
\begin{equation}
\delta = \frac{R_{\rm{vac}}}{2g M^2 (b+\log [ 2\cosh b])} \;,
\end{equation}
depending on the other parameters $g$ and $b$. Choosing a correct value of $\d$ is not enough to realize the current accelerated expansion. The theory has to have at least a stable de Sitter solution, which means that we have to choose the model parameters to satisfy the de Sitter condition and the stability condition \cite{Motohashi:2011wy}.
For the existence of the solution, the parameter $g$ and $b$ has to be in the range \cite{Motohashi:2012tt}  
\begin{equation}
\frac{1}{4} + \frac{0.28}{(b-0.46)^{0.81}} \leq g \leq \frac{1}{2} \;. \label{gcon}
\end{equation}

The third term in Eq.~(\ref{eq58}) smoothly connects the accelerated expansion in the early Universe and the following reheating, radiation-dominated, matter-dominated, and current accelerating Universes, without giving rise to instability of the theory. As a consequence, the third term significantly alters reheating dynamics after the inflation. In the original $R^2$ model, the scalaron oscillates harmonically and reheats the Universe. However, in the presence of the third term, the scalaron oscillates anharmonically.

\begin{figure}[h]
\centering
\includegraphics[width=85mm]{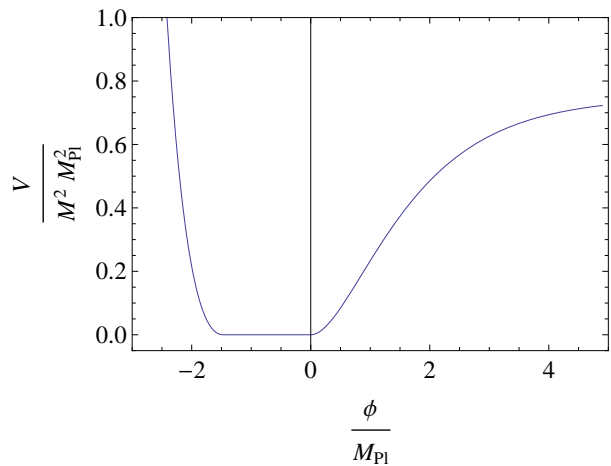}
\caption{
Inflaton potential of the $gR^2$-AB model in the Einstein frame for parameters $g=0.35$, $b=5$, and $\d=5\times 10^{-8}$. For $\phi>0$ and $\phi<\sqrt{6}\Mpl\log \gamma$, the potential is similar to that in pure $R^2$ inflation. On the other hand, for $\sqrt{6}\Mpl\log \gamma<\phi<0$, there is the characteristic plateau. The scalaron starts slow rolling from $\phi>0$, and enters the plateau with large kinetic energy and oscillates inside it.}
\label{fig:pot}
\end{figure}

Since the equations of motion in the Jordan frame is complicated, it would be helpful to consider in the Einstein frame in order to intuitively understand the dynamics of the inflation and reheating. Scalar-field potential in the Einstein frame is shown in Fig.~\ref{fig:pot}. The scalar field starts slow rolling from $\phi>0$, plays a role of the inflaton, and drives quasi-de Sitter expansion. This is also true for the scale factor in the Jordan frame because in Eq.~(\ref{ta}) the scale factors in both frames are related by multiplying an exponential factor of $\phi$, which is almost constant during slow-roll regime. For $\phi>0$, the potential coincides with that of a $R^2$ model and is almost independent of the model parameters $g,~b$ and $\d$. As the scalaron approaches $\phi=0$, it rolls faster and enters the potential plateau with the kinetic energy larger than the potential energy. Then the scalaron oscillates in the plateau and gradually loses its kinetic energy. During the oscillation, the scale factor in the Jordan frame undergoes the periodic evolution due to the exponential factor in Eq.~(\ref{ta}) \cite{Motohashi:2012tt}. At much later time, the scalar field is trapped by the false vacuum and its nonzero potential energy drives the current accelerated expansion of the Universe. 

\subsection{Analytic solutions of $f(R)$ reheating}

To investigate the reheating dynamics, it is easy to solve the motion of the scalar field by working in the Einstein frame and to translate it back to the Jordan frame. In this section, we summarize the basic results that is needed for the later sections of this paper. As for concrete derivation of the solutions, see Ref.~\cite{Motohashi:2012tt}. 

In the inflation and reheating in $f(R)$ gravity, there is no inflaton field from the point of view in the Jordan frame. Consequently, particle creation occurs not through the decay of the inflaton but through the gravitational particle creation~\cite{Starobinsky:1981vz,Vilenkin:1985md,Mijic:1986iv,Ford:1986sy}. Here we use the word "reheating" in the meaning that the energy density of the created radiation is subdominant in the total energy density at the end of inflation but after that it gradually dominates the energy component of the Universe, not inflaton decay. We introduce a massless scalar field $\chi$ into the matter action,
\be S=\int d^4x \sqrt{-g} \kk{\f{\Mpl^2}{2}f(R) -\f{1}{2}g^{\mu\nu} \partial_{\mu} \chi \partial_{\nu} \chi -\f{1}{2} \xi R\chi^2}, \ee
where $\xi$ is a coupling parameter between $\chi$ and gravity. 
The number density of the created scalar particles is~\cite{Starobinsky:1981vz,Vilenkin:1985md,Mijic:1986iv},
\be n(t)=\f{(1-6\xi)^2}{576\pi a^3} \int_{-\infty}^{t} dt^{\prime} a^3R^2, \label{eq60}  \ee
which holds regardless of the functional form of $f(R)$. 

For $\phi>0$ the potential is reduced to that of the pure $R^2$ model, for which the slow-roll approximation can be implemented. Then the field equations \eqref{Eeq1} - \eqref{Eeq3} are analytically solved and the initial conditions for reheating era are obtained \cite{Motohashi:2012tt}. However, the slow-roll approximation does not hold at the transition stage from the slow-roll to the fast-roll. To estimate reheating temperature precisely, we need to use accurate boundary conditions. From the numerical calculation performed in \cite{Motohashi:2012tt}, we found $H_{\rm{end}} =0.26\,M$, where the subscript "end" denotes quantities at the time $t=t_{\rm{end}}$ when the scalar field $\phi$ first crosses zero. Also in the $R^2$ model, the integral in Eq.~(\ref{eq60}) can be performed analytically by substituting the analytic solutions of inflation under slow-roll approximation in the Einstein frame. According to \cite{Motohashi:2012tt}, the energy density of created radiation at $t=t_{\rm{end}}$ in the Jordan frame turns out to be
\begin{equation}
\rho_{r,\rm{end}} = \f{c_0 g_* (1-6\xi)^2}{1152\pi}M^4 \;,
\label{eq25}
\end{equation}
where $g_*$ denotes the relativistic degree of freedom relevant for the particle creation and the constant $c_0$ found in numerical computation is $c_0=0.72$. 

During both inflation and reheating, the energy density of the created radiation is of course subdominant in comparison with that of the inflaton before the reheating completes. Thus, we can neglect its backreaction to the background dynamics. Also after the end of the inflation, the kinetic energy of the inflaton is dominant. Then we can use the fast-roll approximation neglecting potential contribution during the oscillation phase. 

Under these two approximations, we can solve Eqs.~(\ref{Eeq1}) - (\ref{Eeq3}) by separately considering the time intervals dependent on the direction of the motion of the inflaton. To do so, we regard the reflection occurs instantly at potential walls and define the first reflection time of the scalar field at the left wall as $\tilde{t}=\tilde{t}_1$. After that, the inflaton reaches the right wall at $\phi=0$ and is reflected at $\tilde{t}=\tilde{t}_2$. As well, we can periodically define $\tilde{t}_n$. According to \cite{Motohashi:2012tt}, analytic solutions for $\tilde{t}_{n-1}< \tilde{t} < \tilde{t}_n$ in the Einstein frame are as follows:
\begin{align}
\tilde{H}(\tilde{t})&=\f{\tilde{H}_{\rm{end}}}{3\tilde{H}_{\rm{end}}(\tilde{t}-\tilde{t}_{\rm{end}})+1}, 
\label{eq5} \\
\tilde{a}(\tilde{t})&=\tilde{a}_{\rm{end}} \kk{3\tilde{H}_{\rm{end}}(\tilde{t}-\tilde{t}_{\rm{end}})+1}^{1/3}, 
\label{eq6} \\
\frac{\phi(\tilde{t})}{M_{\rm{Pl}}}&=\left\{
\begin{array}{ll}
\displaystyle -\sqrt{\f{2}{3}}\log\kk{3\tilde{H}_{\rm{end}}(\tilde{t}-\tilde{t}_{\rm{end}})+1}-(n-1)\sqrt{6}\log\gamma &\quad (n:{\rm odd}), \\
\\
\displaystyle \sqrt{\f{2}{3}}\log\kk{3\tilde{H}_{\rm{end}}(\tilde{t}-\tilde{t}_{\rm{end}})+1}+n\sqrt{6}\log\gamma &\quad (n:{\rm even}),
\end{array}
\right. 
\label{phifrn} \\
\frac{d}{d\tilde{t}}\phi(\tilde{t})&=\left\{
\begin{array}{ll}
\displaystyle -\sqrt{6}\tilde{H}(\tilde{t}) &\quad (n:{\rm odd}), \\
\\
\displaystyle \sqrt{6}\tilde{H}(\tilde{t}) &\quad (n:{\rm even}), 
\end{array}
\right. \label{dphifrn}
\end{align}
with $\gamma \equiv \sqrt{1-2g}$. Times in the Jordan frame and the Einstein frame are related by
\be 
t(\tilde{t})=
\left\{
\begin{array}{ll}
\displaystyle t_{n-1}+\f{\gamma^{n-1}}{4\tilde{H}_{\rm{end}}} \left[ \left\{ 3 \tilde{H}_{\rm{end}}(\tilde{t}-\tilde{t}_{\rm{end}})+1\right\}^{4/3}-\gamma^{-4(n-1)} \right] &\quad (n:{\rm odd}), \\ 
\\
\displaystyle t_{n-1}+\f{\gamma^{-n}}{2\tilde{H}_{\rm{end}}} \left[ \left\{ 3 \tilde{H}_{\rm{end}}(\tilde{t}-\tilde{t}_{\rm{end}})+1\right\}^{2/3}-\gamma^{-2(n-1)} \right] &\quad (n:{\rm even}),
\end{array}
\right. \label{tJtEfr}
\ee
where $t_{n}$ is given by
\be
t_{n}=
\left\{
\begin{array}{ll}
\displaystyle t_{\rm{end}}+\f{(\gamma^4+\gamma^2+2)(\gamma^{-3(n-1)}-1)}{4\tilde{H}_{\rm{end}}(\gamma^4+\gamma^2+1)}+\f{\gamma^{-4}-1}{4\tilde{H}_{\rm{end}}\gamma^{3(n-1)}} &\quad (n:{\rm odd}), \\
\\
\displaystyle t_{\rm{end}}+\f{(\gamma^4+\gamma^2+2)(\gamma^{-3n}-1)}{4\tilde{H}_{\rm{end}}(\gamma^4+\gamma^2+1)} &\quad (n:{\rm even}).
\end{array}
\right. 
\label{eq4}
\ee
From Eqs.~\eqref{ta}, \eqref{HJ}, \eqref{eq5}, \eqref{eq6}, and \eqref{tJtEfr}, the Hubble parameter and the scale factor in the Jordan frame evolve as
\bea
H(t)&=&
\left\{
\begin{array}{ll}
\displaystyle \f{2\gamma^{3(n-1)}\tilde{H}_{\rm{end}}}{4\gamma^{3(n-1)}\tilde{H}_{\rm{end}}(t-t_{n-1})+1} &\quad (n:{\rm odd}), \\
\\
\displaystyle 0 &\quad (n:{\rm even}),
\end{array}
\right. \label{HJfr} \\
a(t)&=&
\left\{
\begin{array}{ll}
\displaystyle a_{\rm{end}}\gamma^{-(n-1)}\kk{4\gamma^{3(n-1)}\tilde{H}_{\rm{end}}(t-t_{n-1})+1}^{1/2} &\quad (n:{\rm odd}), \\
\\
\displaystyle a_{\rm{end}}\gamma^{-n} &\quad (n:{\rm even}).
\end{array}
\right. 
\eea
$H(t)$ periodically oscillates, jumping at $t=t_{n}$ ($n=1,2,\cdots$) between $H_n=2\tilde{H}_n=\gamma^{3n} H_{\rm{end}}$ and $H_n=0$. 

As discussed in \cite{Motohashi:2012tt}, the time-averaged behavior of the Hubble parameter and scale factor in the Jordan frame are obtained by being careful about inhomogeneous tick of the Jordan-frame time viewed from the Einstein frame and weighting the odd- and even-period by appropriate coefficients. Hence, the averaged Hubble parameter and scale factor are 
\begin{align}
\langle H(t) \rangle = \frac{H_{\rm{end}}}{3 H_{\rm{end}} (t-t_{\rm{end}}) +1} \nonumber \;, \\
\langle a(t) \rangle = a_{\rm{end}} \left[ 3 H_{\rm{end}} (t-t_{\rm{end}}) + 1 \right]^{1/3} \;. \nonumber
\end{align}
By comparing the energy densities of radiation and gravity, the reheating temperature is estimated. The particle creation during the plateau oscillation phase is negligible because $R\simeq b\delta M^2\ll 1$. Therefore $\rho_{\rm{r}}$ approximately scales as $\langle \rho_{\rm{r}}(t) \rangle \propto  \langle a(t) \rangle^{-4} \propto (t-t_{\rm{end}})^{-4/3}$. On the other hand, the effective energy density of gravity is defined with the equation of motion in the Jordan frame as $H^2 = (\rho_r+\rho_{\rm{g}})/(3M_{\rm{Pl}}^2)$. Then we find that the effective energy density of gravity scales as $\langle \rho_{\rm{g}} \rangle \propto (t-t_{\rm{end}})^{-2}$. Therefore, the reheating temperature $T_{\rm{reh}}$ defined by the condition $\langle \rho_{\rm{r}} \rangle =\langle \rho_{\rm{g}} \rangle$ is given by
\begin{equation}
\frac{T_{\rm{reh}}}{M} \approx 8.1 \times 10^{-3} [g_* (1-6\xi)^2]^{1/2} \left( \frac{M}{M_{\rm{Pl}}} \right) \;,
\label{eq12}
\end{equation}
where the boundary conditions at the end of inflation, $c_0=0.72$ and $H_{\rm{end}}\approx 0.26 M$, are used. For minimally coupled scalar fields with $g_*=100$ and $M/\Mpl=1.2\times 10^{-5}$, the reheating temperature is $T_{\rm{reh}} \approx 3.0 \times 10^7\,{\rm{GeV}}$. We emphasize that the parameters $g$, $b$ and $\delta$ considerably alter the dynamics of the reheating in the $gR^2$-AB model but does not affect the reheating temperature, the averaged Hubble parameter, and the averaged scale factor. As we investigate in the next section, crucial parameters for a GWB spectrum are $M$ and the combination $g_* (1-6\xi)^2$. However, $M$ has already been fixed by the observation of CMB. Therefore, $g_* (1-6\xi)^2$ is the only free parameter of $gR^2$-AB model for a GWB spectrum. In the following of this paper, we will show that $g_* (1-6\xi)^2$ is constrained from the current observational limit on the energy abundance of GWs. 

\section{Gravitational waves}
\label{secGW}
We consider the production and evolution of gravitational waves during inflation and the following reheating era. Tensor perturbations $h_{ij}$ have two polarizations and can be written with polarization tensors $e_{ij}^+$ and $e_{ij}^{\times}$ as
\begin{equation}
h_{ij}=h_{+} e_{ij}^{+} +h_{\times} e_{ij}^{\times} \;.
\end{equation}
Hereafter we will work in the transverse-traceless gauge. The polarization tensors are $e_{xx}=-e_{yy}=1$ and $e_{xy}=e_{yx}=1$ for the GW propagating in the $z$ direction. In the Friedmann-Robertson-Walker (FRW) background, each tensor perturbation obeys the following equation \cite{Hwang:1996xh}:
\begin{equation}
\ddot{h}+3H_{\rm{eff}} \dot{h} + \frac{k^2}{a^2} h =0\;,
\label{eq15} 
\end{equation}
where we abbreviate the subscript of the polarizations. The dot denotes the derivative with respect to $t$. The effective Hubble parameter including modification of gravity from GR is defined by 
\begin{equation}
H_{\rm{eff}} \equiv H+\frac{\dot{F}}{3F}\;,
\end{equation}
in which $F$ is given in Eq.~(\ref{FRJ}). Introducing new variables $a_{\rm{eff}} \equiv a \sqrt{F}$ and $u \equiv a_{\rm{eff}} h$, we write Eq.~(\ref{eq15}) as   
\begin{equation}
u^{\prime \prime} + \left( k^2 - \frac{a_{\rm{eff}}^{\prime \prime}}{a_{\rm{eff}}} \right) u =0 \;,
\label{eq17}
\end{equation}
where the prime is the derivative with respect to the conformal time $\tau\equiv \int \f{dt}{a(t)}$.

\subsection{Cosmic expansion}

To solve Eq.~(\ref{eq17}), we need the time evolution of the cosmic background expansion and $\dot{F}/F$ on the FRW background. During inflation era, the $gR^2$-AB model is well approximated by $R^2$ inflation model, which has been well studied.
The deviation from exact de Sitter inflation is parameterized by $\nu$, which is $3/2$ for the de Sitter inflation, and the inflationary expansion is given by  
\begin{equation}
\frac{a_{\rm{eff}}^{\prime \prime}}{a_{\rm{eff}}} = \frac{\nu^2-1/4}{\tau^2} \;. \nonumber
\end{equation} 
In a general model of $f(R)$ gravity, the parameter $\nu$ is related to the inflationary slow-roll parameters \cite{Hwang:2001pu}
\begin{equation}
\epsilon_1 \equiv -\frac{\dot{H}}{H^2} \;, \quad \quad \epsilon_3 \equiv \frac{\dot{F}}{2HF} \;,
\end{equation}
as
\begin{align}
\nu &= \sqrt{\frac{1}{4} + \frac{(1+\epsilon_3)(2-\epsilon_1+\epsilon_3)}{(1-\epsilon_1)^2}} \nonumber \\
&\approx \frac{3}{2} +\epsilon_1+\epsilon_3 \;,
\end{align}
to the first order in the slow-roll parameters. Since the inflation in the $gR^2$-AB model is well approximated by $R^2$ inflation, in which $\epsilon_1 \approx -\epsilon_3$ gives $\nu \approx 3/2$, the inflation is nearly de-Sitter expansion.

During the following cosmic eras, since finding the solutions of the cosmic expansion in the Jordan frame is so complicated, we instead use the solution of a scalar-field motion in the Einstein frame. From Eqs.~(\ref{FRJ}) and (\ref{ta}), 
\begin{equation}
\frac{\dot{F}}{F}=\sqrt{\frac{2}{3}} \frac{\dot{\phi}}{M_{\rm{Pl}}} = \frac{1}{M_{\rm{Pl}}}\sqrt{\frac{2}{3}} \frac{d\phi}{d\tilde{t}} e^{\frac{\phi}{\sqrt{6}M_{\rm{Pl}}}} \;.
\end{equation}

First let us consider the reheating era. During the reheating, the scalar field undergoes anharmonic oscillation. Correspondingly the Hubble parameter in the Jordan frame periodically change its magnitude between $H\approx 1/(2t)$ for $\dot{\phi} <0$ and $H\approx 0$ for $\dot{\phi} >0$ as in Eq.~(\ref{HJfr}). Using the analytic solutions in Eqs.~(\ref{eq5}), (\ref{phifrn}), (\ref{dphifrn}), (\ref{tJtEfr}), and (\ref{HJfr}), we can show $\dot{F}/F=-H$ for $\dot{\phi} <0$. On the other hand, for $\dot{\phi} >0$, using the same set of equations, we have 
\begin{equation}
\frac{\dot{F}}{F}=\frac{2\tilde{H}_{\rm{end}} \gamma^{3n-2}}{2\tilde{H}_{\rm{end}} \gamma^{3n-2}(t-t_{n-1})+1} \;. \nonumber
\end{equation}
Then substituting Eq.~(\ref{eq4}) for even $n$ into the above equation and keeping the leading term in $\gamma$ gives
\begin{equation}
\frac{\dot{F}}{F}=\frac{2\tilde{H}_{\rm{end}} \gamma^{3n-2}}{2\tilde{H}_{\rm{end}} \gamma^{3n-2}(t-t_{\rm{end}})+1-\gamma}
\approx \frac{1}{t-t_{\rm{end}}}\;. \nonumber
\end{equation}
At the last equality, we used the fact that the constraint on $g$ in Eq.~(\ref{gcon}) restricts the range of $\gamma$ to $0 \leq \gamma \lesssim 0.7$. Therefore we obtain for both signs of $\dot{\phi}$,
\begin{equation}
H_{\rm{eff}} \approx \frac{1}{3t} \;, \quad \quad {\rm{for}} \; {\rm{reheating}}\;.
\label{Heff}
\end{equation}

Equation (\ref{Heff}) indicates that the Hubble parameter which tensor perturbation feels is the same as that in the case of perfect fluid domination whose equation of state is $w\equiv p/\rho=1$. This cosmological phase is often called the stiff phase (SP) or the kinetic-energy dominant phase in the case of inflation caused by scalar fields \cite{ArmendarizPicon:1999rj,Peebles:1998qn}. 
The cosmic expansion results in the effective scale factor, $a_{\rm{eff}} \propto t^{1/3}$, which can be obtained directly from $a_{\rm{eff}}=a\sqrt{F}$ by averaging over many periods of the Hubble oscillation during the reheating \cite{Motohashi:2012tt}. Note that $H_{\rm{eff}} \neq \dot{a}_{\rm{eff}}/a_{\rm{eff}}$, but both coincides only after time averaging.

After the end of the reheating, the energy of radiation created during the inflation and reheating dominates the Universe. Then as in the standard cosmology the radiation-dominated era (RD) is followed by the matter-dominated era (MD) and the cosmological constant-dominated era ($\Lambda$D). In the $gR^2$-AB model, the late-time accelerated expansion is caused by modification of gravity. Since $R \ll M^2$ in Eq.~(\ref{FgABR2}) after the reheating, we have 
\begin{equation}
\frac{\dot{F}}{F}= \frac{g\, {\rm{sech}}^2 \sigma}{1-g+g \tanh \sigma} \dot{\sigma} \;, \quad \quad \sigma \equiv \frac{R}{M^2 \delta} -b \;.
\label{eq16}
\end{equation}
In RD, $R=0$ gives $\dot{F}/F=0$ and then $H_{\rm{eff}}=H$. In $\Lambda$D, $R$ is constant and also $H_{\rm{eff}}=H$. In MD, the curvature $R$ should be greater than $R_{\rm{vac}} \sim M^2 \delta$ during MD so that MD is followed by $\Lambda$D. Since $\sigma \gg1$, Eq.~(\ref{eq16}) is approximated to 
\begin{equation}
\frac{\dot{F}}{F} \approx 4g e^{-2\sigma} \dot{\sigma} \sim e^{-2\sigma} \sigma H \ll H \;.
\end{equation}
So the contribution of modified gravity to $H_{\rm{eff}}$, namely $\dot{F}/F$, is much smaller than the physical Hubble parameter $H$. Thus we also have $H_{\rm{eff}}=H$ during MD. In summary, the cosmic expansions in RD, MD, and $\Lambda$D phase are the same as those in the standard cosmology, in contrast to the cosmic expansion during the reheating phase characteristic to the $gR^2$-AB model.

The evolution of the effective scale factor including the modification due to $f(R)$ gravity is expressed by matching its value and the first derivative at their transitions as
\begin{align}
a_{\rm{eff}}^{({\rm{i}})} (\tau) &= a_{\rm{end}} \left( \frac{\tau}{\tau_{\rm{end}}} \right)^{(1-2\nu)/2} \;, \hspace{45mm} {\rm{for\;\;Inf}} : -\infty < \tau < \tau_{\rm{end}} \;, 
\label{eq18} \\
a_{\rm{eff}}^{({\rm{s}})} (\tau) &= a_{\rm{end}}\sqrt{2\nu-1}\, \sqrt{\lambda - \frac{\tau}{\tau_{\rm{end}}}} \;, \hspace{40mm} {\rm{for\;\;SP}} : \tau_{\rm{end}} < \tau < \tau_{\rm{reh}}  \;, \\
a_{\rm{eff}}^{({\rm{r}})} (\tau) &= - \frac{a_{\rm{end}}}{2}\sqrt{2\nu-1}\, \frac{\tau+\tau_{\rm{reh}}-2\lambda \tau_{\rm{end}}}{\sqrt{\tau_{\rm{end}}(\lambda \tau_{\rm{end}}-\tau_{\rm{reh}})}} \;, \hspace{22mm} {\rm{for\;\;RD}} : \tau_{\rm{reh}} < \tau < \tau_{\rm{eq}} \;,  \\
a_{\rm{eff}}^{({\rm{m}})} (\tau) &= -\frac{a_{\rm{end}}}{4}\sqrt{2\nu-1}\, \frac{\,\tau^2 +\tau_{\rm{eq}}(\tau_{\rm{eq}}+2 \tau_{\rm{reh}}-4\lambda \tau_{\rm{end}})\,}{\tau_{\rm{eq}}\sqrt{\tau_{\rm{end}} (\lambda \tau_{\rm{end}}-\tau_{\rm{reh}})}} \;, \quad \quad  {\rm{for\;\;MD}} : \tau_{\rm{eq}} < \tau < \tau_0 \;,
\label{eq19}
\end{align}
where $\tau_{\rm{end}}$, $\tau_{\rm{reh}}$, $\tau_{\rm{eq}}$, and $\tau_{\rm{0}}$ are the conformal time at the end of inflation, the end of reheating era, the matter-radiation equality, and present, respectively, and $a_{\rm{end}}$ is the scale factor at $\tau_{\rm{end}}$. We do not take the $\Lambda$D phase into account for simplicity because it hardly affects observational constraint from GWs as we will see later. In these equations, we defined the parameter 
\begin{equation}
\lambda \equiv \frac{2\nu}{2\nu-1} \;,
\end{equation}
to merely simplify the above equations. 

\subsection{GWB energy spectrum}
In the context of quantum field theory in a curved spacetime, GW production can be interpreted as the amplification of vacuum fluctuations by cosmic expansion (gravitational particle creation) and is inevitable consequence of inflation \cite{Abbott:1985cu,Allen:1987bk,Sahni:1990tx}. We quantize linear GWs, $u$, in Eq.~(\ref{eq17}), and write a graviton field as
\begin{equation}
\hat{u} (\tau,\mathbf{x})= M_{\rm{Pl}}^{-1} \int \frac{d^3k}{(2\pi) ^3 \sqrt{ 2k}} \left[ \hat{b}_{\mathbf{k}} \psi(k,\tau) e^{i\mathbf{k} \cdot \mathbf{x}} + \hat{b}^{\dag}_{\mathbf{k}} \psi^{\ast}(k,\tau) e^{-i\mathbf{k} \cdot \mathbf{x}} \right] \;, \nonumber
\end{equation}
Here $\hat{b}_\mathbf{k}$ and $\hat{b}^{\dag}_\mathbf{k}$ are the annihilation and creation operators, respectively. Substituting this into Eq.~(\ref{eq17}), we have an equation for $\psi$,
\begin{equation}
\psi^{\prime \prime} + \left( k^2 - \frac{a_{\rm{eff}}^{\prime \prime}}{a_{\rm{eff}}} \right) \psi =0 \;.
\end{equation}
In each cosmological era described by Eqs.~(\ref{eq18}) - (\ref{eq19}), the solutions are given in terms of the Hankel function:
\begin{align}
\psi_{\rm{i}} (k,\tau) &= \sqrt{\frac{\pi}{4k}} e^{-i \pi(2\nu+1)/4} \sqrt{x} H_{\nu}^{(2)}(x) \;, \hspace{42mm} {\rm{for\;\;Inf}} : -\infty < \tau < \tau_{\rm{end}}\;, \\
\psi_{\rm{s}} (k,\tau) &= \sqrt{\frac{\pi}{4k}} \sqrt{y} \left[ \alpha_{\rm{s}} (k) e^{-i\pi/4} H_0^{(2)} (y) + \beta_{\rm{s}}(k) e^{i\pi/4} H_0^{(1)} (y) \right] \;, \quad \quad {\rm{for\;\;SP}} : \tau_{\rm{end}} < \tau < \tau_{\rm{reh}} \;, \\
\psi_{\rm{r}} (k,\tau) &= \frac{1}{\sqrt{2k}} \left[ \alpha_{\rm{r}} (k) e^{-ix} + \beta_{\rm{r}} (k) e^{ix} \right] \;, \hspace{42mm} {\rm{for\;\;RD}} : \tau_{\rm{reh}} < \tau < \tau_{\rm{eq}} \;, \\
\psi_{\rm{m}} (k,\tau) &\approx -\sqrt{\frac{\pi}{4k}} \sqrt{w} \left[ \alpha_{\rm{m}} (k) H_{3/2}^{(2)}(w) +\beta_{\rm{m}} (k) H_{3/2}^{(1)}(w)  \right] \;, \hspace{16mm}  {\rm{for\;\; MD}} : \tau_{\rm{eq}} < \tau < \tau_0 \;, 
\end{align}
where
\begin{equation}
x \equiv k\tau \;, \quad \quad y\equiv k\left(\tau-\lambda \tau_{\rm{end}} \right) \;, \quad \quad w \equiv k\left[ \tau+(\sqrt{2}-1)\tau_{\rm{eq}} \right] \;.
\label{eq:xyw}
\end{equation}
The mode function during MD is valid only when $\tau_{\rm{eq}} \gg \tau_{\rm{reh}},|\tau_{\rm{end}}|$, which always holds in the standard cosmic expansion history.
Matching the solutions and their first derivative at the transitions, we obtain the Bogoliubov coefficients, whose precise expressions are given in Appendix \ref{Bogoliubov}. Once the Bogoliubov coefficients are obtained, a GWB spectrum can be computed with the formula \cite{Maggiore:1999vm}:
\begin{equation}
h_0^2 \Omega_{\rm{gw}} (f) = \frac{16\pi^2}{3(H_0/h_0)^2 M_{\rm{Pl}}^2} f^4 |\beta|^2 \;, 
\label{eq20} 
\end{equation}
where $H_0$ is the Hubble parameter at present and $h_0$ is that normalized by $100\,{\rm{km}}\,{\rm{s}}^{-1}\,{\rm{Mpc}}^{-1}$. Note that this energy density contains contribution from both plus and cross tensor polarizations. To express the GWB energy spectrum as a function of frequency, one need to convert $x$, $y$, and $w$ in Eq.~(\ref{eq:xyw}) into frequencies as  
\begin{align}
x_{\rm{end}} &\equiv k\tau_{\rm{end}} =-2 \pi f \left( \frac{a_0}{a_{\rm{end}}} \right) \frac{1+(\nu-3/2)}{H_{\rm{end}}} \approx -\frac{f}{f_{\rm{end}}} \;, \quad \quad
y_{\rm{end}} \equiv (1-\lambda) k \tau_{\rm{end}} \approx -\frac{1}{2} k \tau_{\rm{end}}
\label{eq21} \\
x_{\rm{reh}} &\equiv k \tau_{\rm{reh}}= \frac{f}{f_{\rm{reh}}} \;, \quad \quad 
y_{\rm{reh}} \equiv k \left( \tau_{\rm{reh}} -\lambda \tau_{\rm{end}} \right) \approx k \tau_{\rm{reh}} \;, \\
x_{\rm{eq}} &\equiv k \tau_{\rm{eq}} = \frac{f}{f_{\rm{eq}}} \;, \quad \quad w_{\rm{eq}} \equiv \sqrt{2} k \tau_{\rm{eq}} \;. \label{eq22} 
\end{align}
The approximation in Eq.~(\ref{eq21}) is good only if $\nu \approx 3/2$ as in the $gR^2$-AB model. Substituting Eqs.~(\ref{eqb1}), (\ref{eqb2}), and (\ref{eqb3}) into Eq.~(\ref{eq20}) together with Eqs.~(\ref{eq21}) - (\ref{eq22}), we finally obtain the approximated expressions of a GWB spectrum:
\begin{align}
h_0^2 \Omega_{\rm{gw}} (f) &= 1.73\times 10^{-50} \times 2^{2\nu} \Gamma^2(\nu) \left( \frac{f}{1\,\rm{Hz}} \right)^4 \left( \frac{f}{f_{\rm{end}}} \right)^{-2\nu}\;, \hspace{38mm} {\rm{for}}\;\; f_{\rm{reh}} < f < f_{\rm{end}} \;,
\label{eqGWB1} \\
h_0^2 \Omega_{\rm{gw}} (f) &= 1.73\times 10^{-50} \times 2^{2\nu} \Gamma^2(\nu) \left( \frac{f}{1\,\rm{Hz}} \right)^4 \left( \frac{f}{f_{\rm{end}}} \right)^{-2\nu} \left( \frac{f}{f_{\rm{reh}}} \right)^{-1}\;, \hspace{22mm} {\rm{for}}\;\; f_{\rm{eq}} < f < f_{\rm{reh}} \;, \\
h_0^2 \Omega_{\rm{gw}} (f) &= 1.73\times 10^{-50} \times 2^{2\nu} \Gamma^2(\nu) \left( \frac{f}{1\,\rm{Hz}} \right)^4 \left( \frac{f}{f_{\rm{end}}} \right)^{-2\nu} \left( \frac{f}{f_{\rm{reh}}} \right)^{-1} \left( \frac{f}{f_{\rm{eq}}} \right)^{-2}\;, \quad \quad {\rm{for}}\;\; f_{\rm{0}} < f < f_{\rm{eq}} \;. 
\label{eqGWB2}
\end{align} 
Note that this formula is valid for $|\nu-3/2| \ll 1$, namely, quasi-de Sitter inflation. During the inflation, the $gR^2$-AB model is well approximated by $R^2$ inflation, in which the condition $|\nu-3/2| \ll 1$ holds. 

The characteristic frequencies, $f_{\rm{end}}$ and $f_{\rm{reh}}$, are the current frequencies of GWs that exit the horizon at the end of the inflation and at the end of the reheating era, respectively, and depend on $gR^2$-AB model parameters, that is, the energy scale of inflation $M$, the effective number of relativistic particles $g_*$, and the nonminimal coupling constant $\xi$. Using Eqs.~(\ref{eq25}) and (\ref{eq12}) with the boundary conditions at the end of inflation, $c_0=0.72$ and $H_{\rm{end}}\approx 0.26 M$, we find
\begin{align}
f_{\rm{end}} &= \frac{H_{\rm{end}}}{2\pi} \left( \frac{a_{\rm{end}}}{a_0} \right) \approx \frac{M}{8\pi} \left( \frac{a_{\rm{end}}}{a_0} \right) \nonumber \\
&\approx 3.2 \times 10^{10} \left( \frac{g_* (1-6\xi)^2}{106.75} \right)^{-1/4}\;{\rm{Hz}} \;, \label{eq24} \\
f_{\rm{reh}} &= \frac{H_{\rm{reh}}}{2\pi} \left( \frac{a_{\rm{reh}}}{a_0} \right) \nonumber \\
&\approx 5.1 \times 10^{9} \left( \frac{g_* (1-6\xi)^2}{106.75} \right)^{3/4} \left( \frac{M}{M_{\rm{Pl}}} \right)^2 \;{\rm{Hz}} \;.
\end{align}
The frequency of a GW corresponding to the matter-radiation equality is 
\begin{equation}
f_{\rm{eq}} = 1.1 \times 10^{-16} \Omega_m h_0^2 \;{\rm{Hz}}\;,
\end{equation}
for which we use fixed values of standard $\Lambda$CDM cosmology, $\Omega_m=0.3$, $h_0=0.7$, since their uncertainties just slightly shift $f_{\rm{eq}}$ and does not change our conclusion in this paper.  

\begin{figure}[h]
\begin{center}
\includegraphics[width=10.5cm]{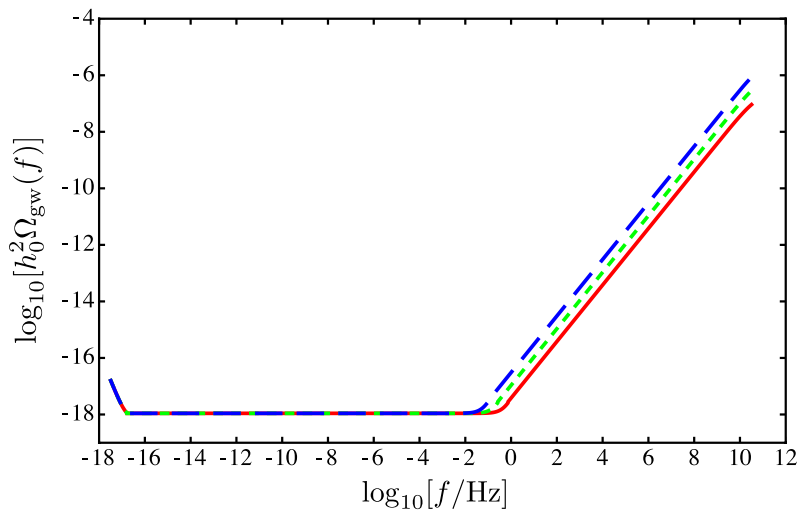}
\caption{GWB spectra for different value of $\xi$ with $M \approx 1.2 \times 10^{-5} M_{\rm{Pl}}$ and $g_*=100$. The lines are $\xi=0$ (red, solid), $\xi=1/12$ (green, dotted), and $\xi=1/8$ (blue, dashed).}
\label{fig19}
\end{center}
\end{figure}

The frequency dependence of the GWB spectrum is the same as that of the standard de Sitter inflationary scenario at low frequencies, but quite different for high frequencies, $f_{\rm{reh}} < f < f_{\rm{end}}$. These high-frequency GWs exit the horizon during the reheating era, whose equation of the state is  $w=1$, and respond to the rapid deceleration of the cosmic expansion. As a result, the GWB spectrum is proportional to $f$ and has a large peak. We see in Eqs.~(\ref{eqGWB1}) - (\ref{eqGWB2}) that the overall amplitude of the GWB spectrum depends only on the parameter $\nu$. In the $gR^2$-AB model, since inflation is caused by $R^2$ term in the action and is nearly de Sitter expansion with $\nu \approx 3/2$, the overall amplitude is fixed. However, the peak amplitude of a GWB spectrum depends on $g_*$, $\xi$, and $M$ through $f_{\rm{end}}$ and $f_{\rm{reh}}$. In Fig.~\ref{fig19}, the GWB spectra for the parameters consistent with observation ($M \approx 1.2 \times 10^{-5} M_{\rm{Pl}}$ and $g_*=100$) are shown, varying the nonminimal coupling parameter, $\xi=0$, $1/12$, and $1/8$. As expected from the dependence of $f_{\rm{reh}}$ on the coupling parameter $\xi$, as the parameter deviates from a conformally coupled case ($\xi=1/6$), $f_{\rm{reh}}$ increases and the amplitude of the GWB spectrum at high frequencies is enhanced. In other words, the less efficient the particle creation is, the longer the reheating lasts, leading to the larger peak of the GWB spectrum.   

In the above calculation, we did not take into account an effect of neutrino free-streaming and time-dependent change of $g_*$ during RD. The free-streaming of relativistic neutrinos, which decoupled from thermal equilibrium at $T \leq 2\,{\rm{MeV}}$, significantly contributes to anisotropic stress, damping the amplitude of a GWB \cite{Weinberg:2003ur}. According to \cite{Watanabe:2006qe}, it has been shown that the neutrino anisotropic stress suppress the amplitude of GWB by 35.5\% in the frequency range between $\approx 10^{-16}\,{\rm{Hz}}$ and $\approx 2\times 10^{-10}\,{\rm{Hz}}$. However, this does not affect our conclusion of this paper because observational constraint from GWs comes from the high-frequency peak of the GWB spectrum. Another effect that we need to consider is the time-dependent change of $g_*$ during RD. $g_*$ is not constant but changes depending on time by responding to the existence of particle species in the Universe. Then it affects the amplitude of a GWB. It is known that $\Omega_{\rm{gw}} (f)$ is corrected by an amount of $[g_*(f)/g_*(f_0)]^{-1/3} \approx 0.32$ if we assume the standard model of particle physics \cite{Watanabe:2006qe}. We will take this suppression of GWB amplitude into consideration when we derive an observational constraint on the $gR^2$-AB model in the next section.

\section{Observational constraints from GWs}
\label{sec:constraint}
We found in the previous section that the GWB spectrum is significantly enhanced at high frequencies. However, planned GW detectors are not enough sensitive to detect the GWB for the parameter $\xi$ chosen in Fig.~\ref{fig19}, because the sensitivity of the ground-based GW detectors under construction such as advanced LIGO, advanced VIRGO, and KAGRA (previously called LCGT) \cite{Losurdo:2012zz} is $h_0^2 \Omega_{\rm{gw}}=10^{-9}$ at $f=100\,{\rm{Hz}}$. More advanced detectors such as Einstein Telescope \cite{Hild:2011np} and BBO/DECIGO \cite{Harry:2006fi,Kawamura:2011zz} can reach $h_0^2 \Omega_{\rm{gw}}=10^{-12}$ at $f=10\,{\rm{Hz}}$ and $h_0^2 \Omega_{\rm{gw}}=10^{-16}$ at $f=0.1\,{\rm{Hz}}$, respectively. But they are not sufficient for the GWB in the $gR^2$-AB model.  

On the other hand, the indirect bound coming from the combination of the observational data of the CMB and the matter power spectrum tightly constrains the energy density of a GWB \cite{Sendra:2012wh}, because the observational limit is imposed on the integral of the energy density over a wide frequency range above the frequency of CMB decoupling ($\sim 10^{-16}\,{\rm{Hz}}$). According to \cite{Sendra:2012wh}, the observational limit is
\begin{equation}
\int_{f_{\rm{min}}}^{f_{\rm{max}}} d(\ln f)\, h_0^2 \Omega_{\rm{gw}}(f) \leq 1.0 \times 10^{-6} \,  \;,
\label{eq23}
\end{equation}      
where $f_{\rm{min}}$ is the frequency corresponding to CMB decoupling, which is $\sim 10^{-16}$\,{\rm{Hz}}, and $f_{\rm{max}}$ is the Planck frequency of $\sim10^{43}\,{\rm{Hz}}$. In practice, $f_{\rm{max}}$ matches the high-frequency cutoff of the GWB spectrum, $f_{\rm{end}}$. The GWB spectrum at high frequencies in Eq.~(\ref{eqGWB1}) is linearly proportional to frequency and can be written as
\begin{equation}
\Omega_{\rm{gw}} (f) \approx \Omega_{\rm{gw}} (f_{\rm{end}}) \left( \frac{f}{f_{\rm{end}}} \right) \;.
\end{equation}
This is predominant contribution for the frequency integral in Eq.~(\ref{eq23}). Then the constraint in Eq.~(\ref{eq23}) is
\begin{equation}
h_0^2 \Omega_{\rm{gw}}(f_{\rm{end}}) \leq 1.0 \times 10^{-6} \,  \;.
\end{equation}

\begin{figure}[h]
\begin{center}
\includegraphics[width=10.5cm]{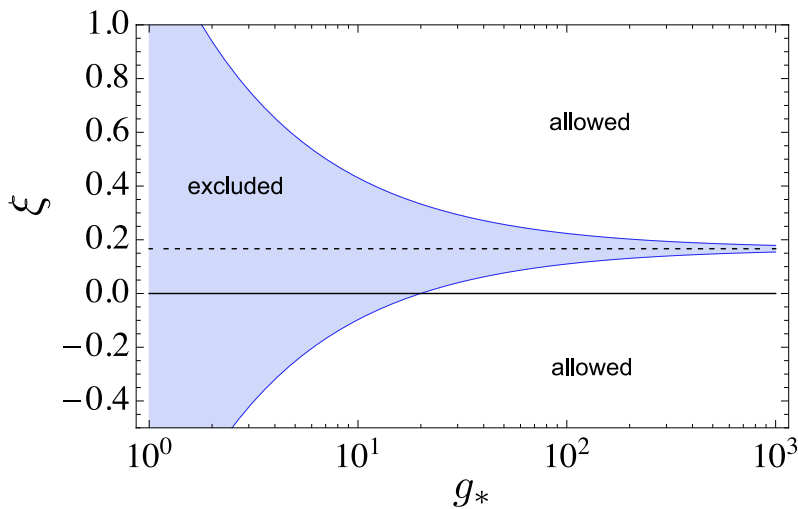}
\caption{Constraint on reheating in the $gR^2$-AB model from a GWB. The shaded region has been excluded by the observations. The horizontal lines represent minimal coupling $\xi=0$ (solid) and conformal coupling $\xi=1/6\approx 0.17$ (dashed), respectively.}
\label{fig1}
\end{center}
\end{figure}

As we mentioned at the end of the previous section, the change of $g_*$ during RD suppresses $\Omega_{\rm{gw}} (f)$ at high-frequencies by a factor of $[g_*(f)/g_*(f_0)]^{-1/3}$ \cite{Watanabe:2006qe}. Taking this fact into account and assuming that $g_*(f)$ is constant for some time duration after the inflation ends, from Eqs.~(\ref{eqGWB1}), (\ref{eq24}), and (\ref{eq23}), we have 
\begin{equation}
g_*^{4/3} (1-6\xi)^2 \geq 54\;, 
\end{equation}
where we used $g_*(f_0) \approx 3.36$ and $\nu \approx 3/2$ for $gR^2$-AB model. In Fig.~\ref{fig1}, the constraint on parameters of reheating, $g_*$ and $\xi$, in the $gR^2$-AB model are shown. For the smaller number of $g_*$, the constraint on the coupling parameter is tighter and the stronger coupling to gravity is required to be compatible with observations. If one assumes that all scalar fields are minimally coupled to gravity ($\xi=0$), many scalar degrees of freedom, at least $g_*=20$, are needed. As the coupling approaches conformal ($\xi=1/6$) case, more scalar fields are necessary . For exactly conformal coupling, no particle creation occurs and the reheating fails to complete. Nevertheless, particle creation could occur at a quantum level via so-called trace anomaly \cite{Birrell&Davies:book,Watanabe:2010vy}. Although more detailed study is needed to reach quantitative conclusion, this effect is of the order of quantum one-loop and is significantly suppressed compared to a tree level. Thus the constraint in Fig.~\ref{fig1} still holds except for the exactly conformal coupling case.

\section{Conclusions and discussion}
\label{sec-cn}
We have studied GW production during the inflation and reheating eras in $f(R)$ gravity, especially in $gR^2$-AB model. In this model, gravity action, which is a function of scalar curvature, is elaborated so as to smoothly connect two accelerated cosmic expansions in the early Universe and at the present time, avoiding instability and singularity in the model. Inflation is described by the original $R^2$ inflation model. However, reheating is quite different from the $R^2$ model because of an additional term in the $f(R)$ action. As a result, the modification of gravity alters cosmic expansion during the reheating phase as if there exists effective fluid with the equation of state of $w=1$. Consequently, a GW spectrum has a large peak at high frequencies. Since the inflation energy scale $M$ is pinned down by the observational data of CMB, the remaining model parameters are the number of degrees of freedom relevant to gravitational particle creation $g_*$ and its coupling constant $\xi$. We have computed the GW spectrum and found that the interesting region of the model parameters has already been excluded by the cosmological limit on abundance of GWs coming from the observational data of the CMB and the matter power spectrum. In the $gR^2$-AB model, the reheating by minimally coupled massless scalar fields require at least $g_*=20$ to be compatible with observations. In the future, the improvement of the sensitivity of the CMB and galaxy survey will provide us more stringent test of inflation and reheating dynamics based on $f(R)$ gravity theory.

Finally, we comment on the result of the recent paper by Kunimitsu and Yokoyama \cite{Kunimitsu:2012xx}. They have shown that if the Higgs field $\varphi$ in the standard model whose mass is $m_h \approx 126\,{\rm{GeV}}$ is minimally coupled to gravity and has positive self-coupling with its magnitude of the order of $\lambda(\mu)\simeq 10^{-2}$ at the energy scale of inflation, the Higgs condensation due to long-wave quantum fluctuations acquired during inflation may significantly contribute to density perturbations during the reheating era. This happens in the model in which reheating occurs through gravitational particle production such as k-inflation \cite{ArmendarizPicon:1999rj} and quintessential inflation \cite{Peebles:1998qn}. As a consequence, the large curvature perturbation generated by the Higgs condensation during the kinetic reheating phase contradicts with the observed amplitude of curvature perturbations today. This indicates that the reheating process due to the gravitational particle production must not last long and then a large peak on a GWB spectrum at high frequency does not exist. 
However, it is nontrivial that this result can be applied to the $gR^2$-AB model. The reason is that, due to the abrupt change in the Hubble parameter in the Jordan frame during the reheating regime, we cannot assume the simple picture that the Higgs field remains constant and starts oscillation at $H\simeq m_{\rm eff}\equiv \sqrt{\lambda \langle \varphi^2 \rangle}$. Since this issue on the Higgs condensation in the $gR^2$-AB model is beyond the scope of the present paper, we leave it for a future work.

\begin{acknowledgments}
We would like to thank Y.~Watanabe and J.~Yokoyama for valuable comments. A.N. and H.M. are supported by Grant-in-Aid for Scientific Research on Innovative Areas, No.24103006. H.M. is also supported by JSPS Postdoctoral Fellowships for Research Abroad.
\end{acknowledgments}

\appendix

\section{Bogoliubov coefficients}
\label{Bogoliubov}

Here the Bogoliubov coefficients when inflation is followed by SP, RD, and MD eras are summarized.

\begin{itemize}
\item{SP ($f_{\rm{reh}} < f < f_{\rm{end}}$)}
\begin{align}
\alpha_{\rm{s}} &= - \frac{\pi}{4\sqrt{2}} e^{i \pi \nu /2} \left[ H_0^{(1)}\left(\frac{|x_{\rm{end}}|}{2} \right) \left\{ x_{\rm{end}} H_{\nu+1}^{(1)}(|x_{\rm{end}}|)+ \left( \nu+\frac{3}{2} \right) H_{\nu}^{(1)}(|x_{\rm{end}}|) \right\} \right. \nonumber \\
&\left. \quad \quad \quad \quad \quad \quad \quad + x_{\rm{end}} H_{1}^{(1)}\left( \frac{|x_{\rm{end}}|}{2} \right) H_{\nu}^{(1)}(|x_{\rm{end}}|) \right] \;, \\
\beta_{\rm{s}} &= -\frac{i \pi}{4\sqrt{2}} e^{i \pi \nu/2 } \left[ H_0^{(2)}\left( \frac{|x_{\rm{end}}|}{2} \right) \left\{ x_{\rm{end}} H_{\nu+1}^{(1)}(|x_{\rm{end}}|)+ \left( \nu+\frac{3}{2} \right) H_{\nu}^{(1)}(|x_{\rm{end}}|) \right\} \right. \nonumber \\
&\left. \quad \quad \quad \quad \quad \quad \quad + x_{\rm{end}} H_{1}^{(2)}\left( \frac{|x_{\rm{end}}|}{2} \right) H_{\nu}^{(1)}(|x_{\rm{end}}|) \right] \;. 
\end{align}
The above equations can be approximated with the small argument limit of the Hankel functions,
\begin{align}
H_{\nu}^{(n)} (z) &\rightarrow \left( \frac{z}{2} \right)^{\nu} \frac{1}{\Gamma(\nu+1)} \mp \frac{i}{\pi} \Gamma(\nu) \left( \frac{z}{2} \right)^{-\nu} \;, \quad \quad \quad {\rm{for}} \;\; \nu \neq 0 \;, \\
H_{0}^{(n)} (z) &\rightarrow 1 \pm \frac{2i}{\pi} \log z  \;, 
\end{align}
where the upper and lower signs corresponds to the Hankel functions of the first and the second kinds, namely, $n=1$ and $2$, respectively. Using these formulas for $|x_{\rm{end}}| \ll 1$, we obtain the relation
\begin{equation}
\alpha_{\rm{s}}, \beta_{\rm{s}}\;\propto \; 2^{\nu} \Gamma (\nu) |x_{\rm{end}}|^{-\nu}\;. \label{eqb1}
\end{equation}

\item{RD ($f_{\rm{eq}} < f < f_{\rm{reh}}$)}
\begin{align}
\alpha_{\rm{r}} &= \frac{e^{ix_{\rm{reh}}}}{2} \sqrt{\frac{\pi}{2y_{\rm{reh}}}} e^{-i \pi/4} \left[  \alpha_{\rm{s}} \left\{ \left( y_{\rm{reh}} +\frac{i}{2}\right) H_0^{(2)} (y_{\rm{reh}})-i y_{\rm{reh}} H_1^{(2)} (y_{\rm{reh}}) \right\} \right. \nonumber \\
& \left. \quad \quad \quad \quad \quad \quad \quad \quad \quad \; +i \beta_{\rm{s}} \left\{ \left( y_{\rm{reh}} +\frac{i}{2} \right) H_0^{(1)} (y_{\rm{reh}})-i y_{\rm{reh}} H_1^{(1)} (y_{\rm{reh}}) \right\} \right] \;, \\
\beta_{\rm{r}} &= \frac{e^{-ix_{\rm{reh}}}}{2} \sqrt{\frac{\pi}{2y_{\rm{reh}}}} e^{-i \pi/4} \left[ \alpha_{\rm{s}} \left\{ \left( y_{\rm{reh}} -\frac{i}{2}\right) H_0^{(2)} (y_{\rm{reh}})+i y_{\rm{reh}} H_1^{(2)} (y_{\rm{reh}}) \right\} \right. \nonumber \\
& \left. \quad \quad \quad \quad \quad \quad \quad \quad \quad \quad + i \beta_{\rm{s}} \left\{ \left( y_{\rm{reh}} -\frac{i}{2} \right) H_0^{(1)} (y_{\rm{reh}})+i y_{\rm{reh}} H_1^{(1)} (y_{\rm{reh}}) \right\} \right] \;. 
\end{align}
For $y_{\rm{reh}} \ll 1$,
\begin{equation}
\alpha_{\rm{r}}, \beta_{\rm{r}}\;\propto \; 2^{\nu} \Gamma (\nu) |x_{\rm{end}}|^{-\nu} y_{\rm{reh}}^{-1/2}\;. \label{eqb2}
\end{equation}

\item{MD ($f_0 < f < f_{\rm{eq}}$)}
\begin{align}
\alpha_{\rm{m}} &= -\frac{1}{2} \sqrt{\frac{\pi}{2 w_{\rm{eq}}}} \left[ \alpha_{\rm{r}} e^{-i x_{\rm{eq}}} \left\{ (w_{\rm{eq}}-2i) H_{3/2}^{(1)} (w_{\rm{eq}}) +i w_{\rm{eq}} H_{5/2}^{(1)} (w_{\rm{eq}}) \right\} \right. \nonumber \\
& \quad \quad \quad \quad \quad \quad  \left. -\beta_{\rm{r}} e^{i x_{\rm{eq}}} \left\{ (w_{\rm{eq}}+2i) H_{3/2}^{(1)} (w_{\rm{eq}}) - i w_{\rm{eq}} H_{5/2}^{(1)} (w_{\rm{eq}}) \right\} \right] \;, \\
\beta_{\rm{m}} &= \frac{1}{2} \sqrt{\frac{\pi}{2 w_{\rm{eq}}}} \left[ \alpha_{\rm{r}} e^{-i x_{\rm{eq}}} \left\{ (w_{\rm{eq}}-2i) H_{3/2}^{(2)} (w_{\rm{eq}}) +i w_{\rm{eq}} H_{5/2}^{(2)} (w_{\rm{eq}}) \right\} \right. \nonumber \\
& \quad \quad \quad \quad \quad \left. -\beta_{\rm{r}} e^{i x_{\rm{eq}}} \left\{ (w_{\rm{eq}}+2i) H_{3/2}^{(2)} (w_{\rm{eq}}) - i w_{\rm{eq}} H_{5/2}^{(2)} (w_{\rm{eq}}) \right\} \right] \;,
\end{align}
For $w_{\rm{eq}} \ll 1$,
\begin{equation}
\alpha_{\rm{m}}, \beta_{\rm{m}}\;\propto \; 2^{\nu} \Gamma (\nu) |x_{\rm{end}}|^{-\nu} y_{\rm{reh}}^{-1/2} x_{\rm{eq}}^{-1} \;. \label{eqb3}
\end{equation}

\end{itemize}

\bibliography{/Volumes/USB-MEMORY/my-research/bibliography}
\end{document}